\documentclass[conference]{IEEEtran}
\usepackage{times}

\usepackage[noadjust]{cite}
\usepackage[dvips]{graphicx}
\usepackage{psfrag}
\usepackage{subfigure}
\usepackage{amsmath}
\usepackage{algorithmic,algorithm,indentfirst}
\usepackage{hyperref}
\usepackage{units}
\usepackage{color}
\usepackage{soul}
\usepackage[nolist]{acronym}
\usepackage{epsfig}
\usepackage{epstopdf}
\usepackage{booktabs}
\usepackage{array}

\begin{document}

\bstctlcite{IEEEexample:BSTcontrol}

\title{Simulating Dense Small Cell Networks}

\author{Pedro Alvarez, Carlo Galiotto, Jonathan van de Belt, Danny Finn, Hamed~Ahmadi, Luiz DaSilva\\
CONNECT - The Centre for Future Networks and Communications, Trinity College, Dublin, Ireland \\  \{pinheirp, galiotc, vandebej, finnda, ahmadih, dasilval\}@tcd.ie.
\vspace{-5mm}
}

\maketitle

\begin{abstract}

Through massive deployment of additional small cell infrastructure, \acp{dsn} are expected to help meet the foreseen
increase in traffic demand on cellular networks. Performance assessment of architectural and protocol solutions tailored to \acp{dsn} will
require system and network level simulators that can appropriately model the complex interference environment found in those networks. This
paper identifies the main features of \ac{dsn} simulators, and guides the reader in the selection of
an appropriate simulator for their desired investigations. We extend our discussion with a comparison of representative \ac{dsn} simulators. %
\vspace{-0.2cm}

\end{abstract}

\IEEEpeerreviewmaketitle


\begin{acronym}

\acro{3gpp}[3GPP]{3\textsuperscript{rd} Generation Partnership Program}
\acro{arq}[ARQ]{Automatic Repeat ReQuest}
\acro{awgn}[AWGN]{Additive White Gaussian Noise}
\acro{bcqi}[B-CQI]{Best Channel Quality Indicator}
\acro{bler}[BLER]{BLock Error Rate}
\acro{ca}[CA]{Carrier Aggregation}
\acro{comp}[CoMP]{Coordinated Multi-Point transmission and reception}
\acro{cqi}[CQI]{Channel Quality Indicator}
\acro{csi}[CSI]{Channel State Information}
\acro{dr}[DR]{Deployment Ratio}
\acro{dsl}[DSL]{Digital Subscriber Line}
\acro{dsn}[DSN]{Dense Small cell Network}
\acro{eesm}[EESM]{Exponential Effective \ac{sinr} Mapping}
\acro{enb}[eNB]{evolved Node Base station} 
\acro{epc}[EPC]{Evolved Packet Core}
\acro{ffr}[FFR]{Fractional Frequency Reuse}
\acro{ftp}[FTP]{File Transfer Protocol}
\acro{harq}[HARQ]{Hybrid Automatic Repeat reQuest}
\acro{hetnets}[HetNets]{Heterogeneous Networks}
\acro{icic}[ICIC]{Inter Cell Interference Coordination}
\acro{imt-a}[IMT-A]{International Mobile Telecommunications-Advanced}
\acro{irc}[IRC]{Interference Rejection Combining}
\acro{l2s}[L2S]{Link-to-System}
\acro{ll}[LL]{Link Level}
\acro{lte}[LTE]{Long Term Evolution}
\acro{lte-a}[LTE-A]{LTE-Advanced}
\acro{mac}[MAC]{Medium Access Control}
\acro{mcs}[MCS]{Modulation and Coding Scheme}
\acro{miesm}[MIESM]{Mutual Information Effective \ac{sinr} Mapping}
\acro{mimo}[MIMO]{Multiple Input Multiple Output}
\acro{ml}[ML]{Maximum Likelihood}
\acro{mmse}[MMSE]{Minimum Mean Squared Error}
\acro{mrc}[MRC]{Maximum Ratio Combining}
\acro{mu-mimo}[MU-MIMO]{Multi-User \ac{mimo}}
\acro{mui}[MUI]{Multi-User Interference}
\acro{nas}[NAS]{Non-Access Stratum}
\acro{nl}[NL]{Network Level}
\acro{oop}[OOP]{Object Oriented Programing}
\acro{pdcp}[PDCP]{Packet Data Convergence Protocol}
\acro{pf}[PF]{Proportional Fair}
\acro{phy}[PHY]{Physical layer} 
\acro{pmi}[PMI]{Precoding Matrix Indicator}
\acro{pon}[PON]{Passive Optical Network}
\acro{qos}[QoS]{Quality-of-Service}
\acro{ran}[RAN]{Radio Access Network}
\acro{rb}[RB]{Resource Block}
\acro{rlc}[RLC]{Radio Link Control}
\acro{rr}[RR]{Round Robin}
\acro{rrc}[RRC]{Radio Resource Control}
\acro{rrh}[RRH]{Remote Radio Head}
\acro{rrm}[RRM]{Radio Resource Management}
\acro{sdr}[SDR]{Software Defined Networks}
\acro{snr}[SNR]{Signal to Noise Ratio}
\acro{sinr}[SINR]{Signal to Interference and Noise Ratio}
\acro{siso}[SISO]{Single Input Single Output}
\acro{sl}[SL]{System Level}
\acro{sic}[SIC]{Successive Interference Cancellation}
\acro{su-mimo}[SU-MIMO]{Single-User \ac{mimo}}
\acro{tbs}[TBS]{Transport Block Size}
\acro{ts}[TS]{Technical Specification}
\acro{tti}[TTI]{Transmission Time Interval}
\acro{ue}[UE]{User Equipment}
\acro{voip}[VoIP]{Voice over IP}
\acro{xgpon}[XG-PON]{10-Gigabit-capable Passive Optical Network}
\acro{zf}[ZF]{Zero-Forcing}

\end{acronym}


\section{Introduction}

In modern mobile networks, the accepted definition of Heterogeneous Networks (HetNet) has gradually evolved toward extreme network
densification. Researchers, equipment vendors, and network operators are counting on the massive deployment of small cells as a key coping
strategy for the foreseen data tsunami. The high throughput gain offered by the addition of these low-power base stations is hoped to
provide a solution that is efficient in terms of energy, spectrum and cost. Nonetheless, technological advances always go hand in hand with
new technical challenges.

\begin{figure}
    \centering  \includegraphics[width=0.9\columnwidth]{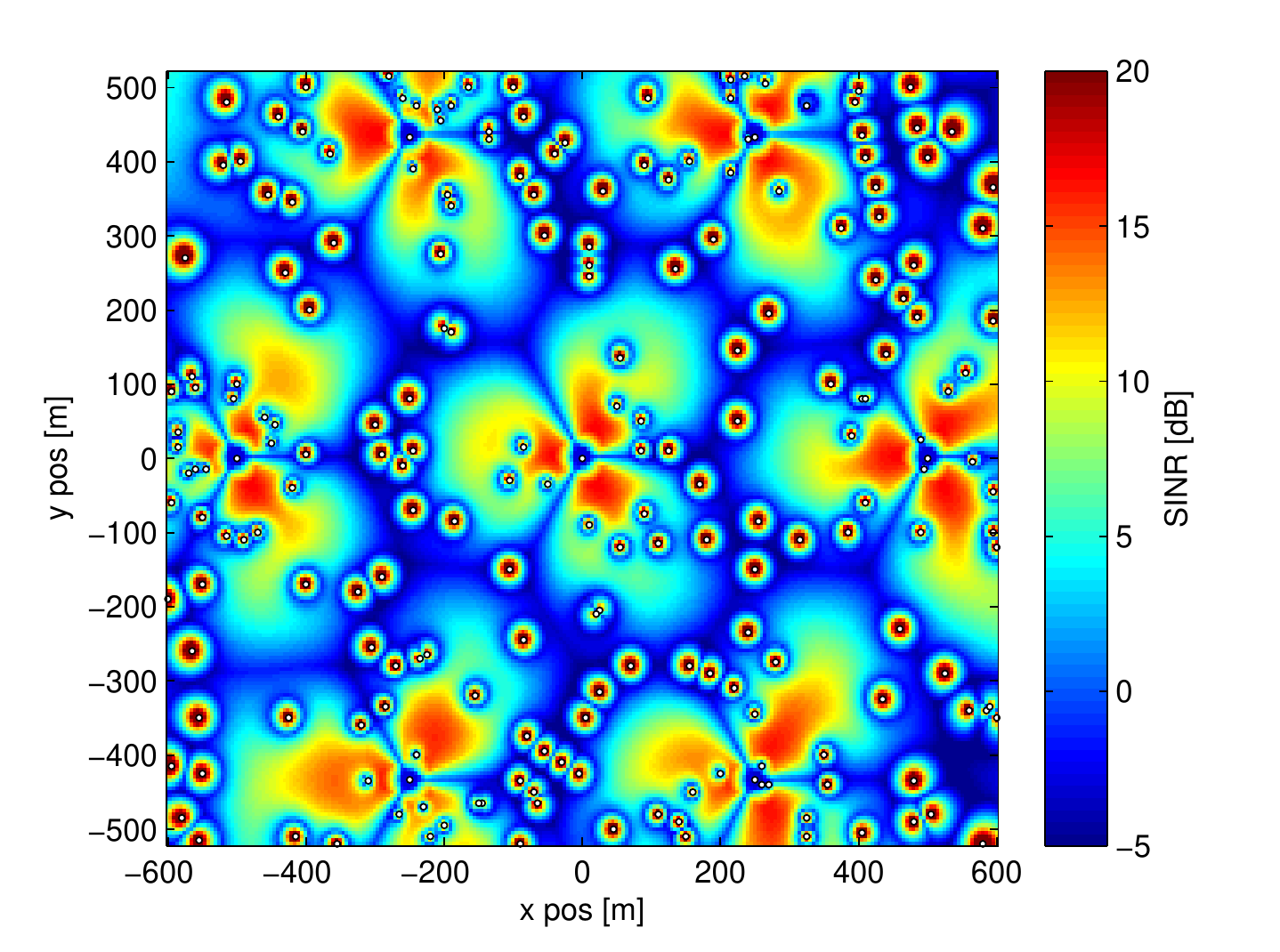}
    \caption{\footnotesize SINR distribution in a dense heterogeneous deployment.}
    \label{fig:edn}
    \vspace{-5mm}
\end{figure}

As illustrated in Figure \ref{fig:edn}, \acfp{dsn} are characterized by massive small cell deployments which enable the operator to offload
low-mobility user traffic away from the macrocell network. Despite the clear advantages of these networks, their high density can
potentially result in high inter-cell interference, bottlenecks in the backhaul and increased energy consumption. To address these
challenges, international and European research projects have begun exploring flexible network architecture designs and user-cell
association procedures.

Due to the scale of these scenarios, simulation is an essential tool for investigating potential \ac{dsn} solutions. 
The main contribution of this paper is in identifying and assessing the tradeoffs among the main simulation approaches for the performance assessment of \acp{dsn}.
Our considerations include, firstly, the choice of the best starting point, whether it be commencing from scratch or extending existing simulators (and if so, which ones?) taking into account the investigation requirements; secondly, supported features, e.g. implemented backhaul protocols
or support for interference suppressing receivers; and thirdly, practical considerations such as ease-of-use, extensibility and run time.

We compare a representative set of \ac{dsn} simulators looking at popular openly available system- and  network-level simulators, as well
as an in-house-built system level simulator with reduced functionality but more specific \ac{dsn} focus. 

\section{Simulator Types}

\begin{figure}
    \centering
    \includegraphics[width=0.8\columnwidth]{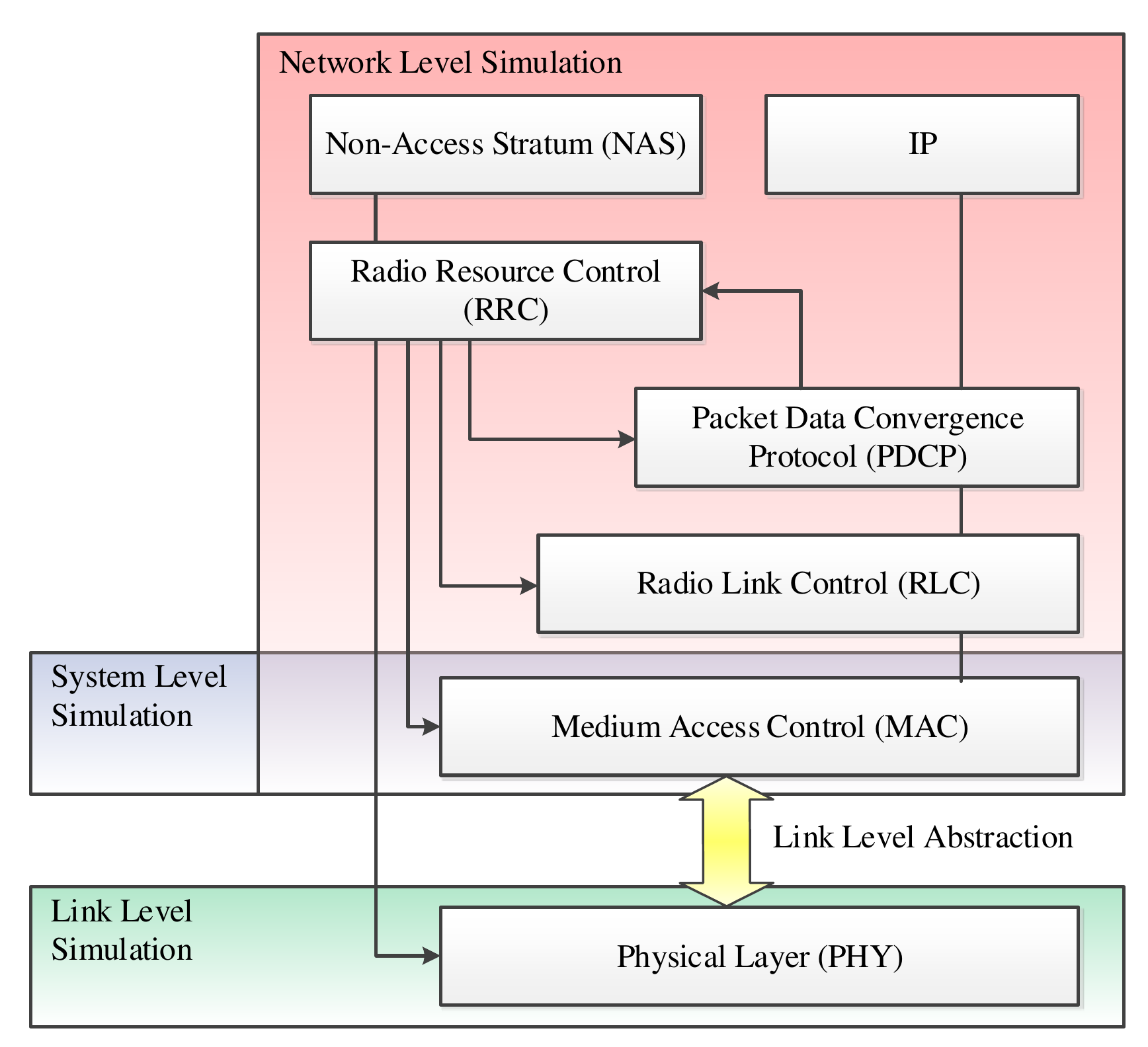}
    \caption{Mapping between simulator types and the layers of the LTE protocol stack. The groupings shown represent our general view and are not necessarily the case for \textit{all} simulator implementations or investigations.}
	\vspace{-0.4cm}
    \label{fig:protocol_stack}
\end{figure}

\ac{dsn} Simulators are generally categorised as either link, system or network level, shown in Figure \ref{fig:protocol_stack}. The implementation of simulators depends on the layer(s) of the protocol stack on which they focus.

\textbf{\ac{ll} simulation} tends to examine the performance of \ac{phy} functionality, usually for a single link between two wireless transceivers. To obtain accurate single-link \ac{bler} statistics, \ac{ll} simulators usually implement all blocks of the \ac{phy} in detail with minimal abstraction. In the \ac{dsn} context, \ac{ll} simulators can be used to compute the \ac{bler} curves for system and network level simulators, which rely on link level abstraction to simulate the \ac{phy}.

When studying complex systems consisting of several base stations and user terminals, \textbf{\acf{sl} simulators} are preferred to \ac{ll}
due to their reduced simulation times. \ac{sl} simulators enable the assessment of coverage, spectral efficiency, and throughput in
wireless networks, taking into account interference generated by neighbouring \acp{enb} or \acp{ue}. \ac{sl} simulators are also suitable
for studying how \ac{rrm} and interference coordination techniques perform and how potential solutions scale with the size of the network.

In contrast to \ac{sl} simulators, which tend to be protocol-agnostic and focus on the air interface, \textbf{\acf{nl} simulators} are designed to
facilitate investigations into specific protocols and their interactions with the upper and lower layers. Moreover, by treating base
stations as network entities capable of exchanging messages between one another, \ac{nl} simulators can model and assess \ac{dsn} backhaul
issues.

\section{Analysis of Desired Features for DSN Simulation}\label{sec:supported_features}

In this section we first outline some important technical features and functionalities necessary for \acl{dsn} simulation. We then examine several operational requirements for using and implementing these in simulators.

\subsection{Technical Features}\label{subsect:technical_features}
There are many technical features that should be implemented to simulate a wide range of \acp{dsn} scenarios with a high degree of accuracy. We consider some of the most important ones below.

\subsubsection{Link Level Abstraction}\label{subsect:Link_level_abstraction}
Full implementation of all links at bit-level would impose a heavy computational burden on system level and network level simulators, especially when the number of links to simulate grows large. To deal with this problem, link level abstraction is used, also known as a \ac{l2s} interface. Subband \acp{sinr} are mapped to a single effective \ac{sinr}, from which the expected \ac{bler} can be computed. Combined with knowledge of the used \ac{mcs} and bandwidth, this allows the determination of the number of successfully transmitted bits within each time frame.
To map the effective \ac{sinr} to the corresponding \ac{bler}, a set of curves (one corresponding to each \ac{mcs}) are obtained from \ac{ll} simulations.

\ac{miesm} is commonly used in \ac{l2s}
interfacing to compute the effective \ac{sinr} because it is known to achieve better performance than most other techniques \cite{Brueninghaus2005}.
Less computationally intensive methods such as the modified Shannon capacity formula \cite{Mogensen2007} also exist for computing \acp{ue}' data rates in \ac{lte} systems. However, these methods often reduce the accuracy  of the \ac{l2s} due to their higher level of abstraction.

The \ac{l2s} interface is of particular importance to \acp{dsn} if one needs to implement interference cancellation or multi-antenna
transmission techniques. As these techniques work at the \ac{phy} level, a more detailed \ac{phy} abstraction provides more accurate simulation results.

\subsubsection{Interference Supressing Receivers}\label{subsect:Interference_Supressing_Receivers}

As the network topology develops toward a more numerous, heterogeneous deployment of nodes, interference becomes a critical issue.

Hence, the implementation of interference cancellation and suppression techniques will be essential for \ac{dsn} applications.

Many linear interference suppressing receivers exist, some focusing on the suppression of interference between different spatial layers (\ac{zf}, \ac{mmse}), others on reducing inter-cell interference from neighboring cells (\ac{irc}).

Several interference suppression techniques make use of multiple receive antennas, combining their signals to isolate the desired signal from interference and noise. Hence, this is closely linked to the \ac{mimo} implementation of the simulator.

\subsubsection{Spectrum Reuse and Inter-cell Interference Coordination}\label{subsubsect:Spectrum_Reuse}

One of the first strategies adopted to avoid interference between adjacent cells was to limit the reuse of spectrum; less interference is caused using higher spectrum reuse factors. However, this implies lower spectral efficiency in the network. Hence, the choice of spectrum reuse factor is subject to a trade-off between mitigating interference and spectral efficiency.

Re-examining spectrum reuse factors may be necessary for \acp{dsn}, as interference among neighbouring cells tends to increase with the density of base stations.
For instance, \ac{lte} has been designed to work with both low reuse factors (e.g., 1) and \ac{ffr} in macrocell scenarios; however, it is likely that the reuse factor for \acp{dsn} will increase as a function of the cell density.
Moreover, given that cells in \acp{dsn} are likely to have an irregular spatial deployment, it is not possible to use standard spectrum reuse patterns developed for hexagonal macrocells grids.
Simulators should allow variation of the spectrum assignment and reuse strategies depending on the cell density and distribution patterns.

There are several other mechanisms in which interference coordination can be carried out. For example, power control strategies and \ac{comp} can be considered as interference coordination techniques.

\subsubsection{Traffic types}\label{subsect:Traffic_types}

Both the volume of traffic and the number of traffic service types is expected to grow hugely in \acp{dsn} where heterogeneous devices and
technologies coexist \cite{Auroux}.

If the volume of users does not increase at the same rate as the cell density, the average number of users per base station will tend to decrease. We can then infer that the smaller the size of the cell, the more dynamic the traffic will be, meaning that smaller cells might switch between active (i.e., users to serve) and inactive states frequently. This will cause time variations in the interference generated by the cell. For this reason, \ac{sinr} distributions, which are usually obtained assuming full buffer traffic, will not give a full picture of network performance and should be complemented with other metrics, such as user perceived rate \cite{Andrews2013}.

Hence, simulators should accurately model various traffic types such as video uploading, video streaming, \ac{voip}, machine-type communications, online multiplayer gaming traffic, while also providing the ability to generate new traffic services.

\subsubsection{Scheduling}\label{subsect:Scheduling} Scheduling algorithms play an essential role in optimizing different aspects of network performance, such as cell throughput or \ac{qos} of \acp{ue}.
The choice of simulator to use for scheduler testing will depend on both the metrics to optimize and the types of feedback available for scheduler decisions, e.g. \ac{csi}.

\ac{sl} and \ac{nl} simulators should support schedulers for various traffic types, such as full-buffer and finite-buffer traffic. Common full-buffer scheduling implementations include \ac{rr} and channel-aware scheduling such as \ac{pf}, \ac{bcqi} and Max-Min Throughput. For finite-buffer traffic additional scheduling methods exist, such as channel-and-\ac{qos} aware scheduling and priority set scheduling for users with and without guaranteed bit rate.

\subsubsection{Upper layers}\label{subsect:Upper_layers}

Packets usually go through many protocol layers before reaching their final destination. These layers, due to flow control and/or \ac{arq} mechanisms, will influence the delay and the amount of traffic transmitted through the air interface, which affects the user-perceived performance. For this reason, modeling these mechanisms is important when simulating \acp{dsn}. Moreover, to simulate some aspects of the network, such as handover performance, connection establishment times, performance of backhaul traffic, etc., functionality above the \ac{mac} layer is required.

While system level simulators usually model only up to the \ac{mac} layer, network level simulators allow the user to investigate the system all the way up to the application layer.

\subsubsection{Backhaul}\label{subsect:Backhaul}

In \aclp{dsn}, multiple base stations are connected to the core network through many heterogeneous backhaul technologies such as point-to-point fiber links, \acp{pon}, \ac{dsl}, microwave relays, etc.
Frequently, studies of mobile networks assume an ideal backhaul, in which the fixed links have no delay or bandwidth limitations.
In practice, the capacity and latency of the backhaul are constrained and this can degrade the performance of the wireless network, especially when there is cooperation between base stations.
Usually the design of the backhaul network is made independently from the wireless network.

However, the co-design of the wireless and optical networks in an integrated manner can improve the utilization of both. This is particularly true of DSNs, where, due to the large number of base stations, technologies other than point-to-point links (e.g., \ac{comp}) should also be considered.

To study these issues, \ac{dsn} simulators must be able to model the backhaul and additional transport technologies such as simple point-to-point links or \acp{pon}. Due to the lack of higher layer protocols in system level simulators, network level simulators are better suited for this task.

\subsubsection{Emulation} \label{subsect:Emulation}

Even though simulation can give us valuable insights into the network performance, these do not replace practical prototyping with actual
testbeds. Unfortunately, designing a full system is a complex task, not feasible in many research centers. For this reason, an important
feature a simulator might have is the capability to connect to real hardware, while emulating other parts of the network.

\subsection{Operational Considerations}

In addition to the technical requirements outlined above, \ac{dsn} operational considerations should also be taken into account, with a view towards verification and validation.
In particular, we consider how easily users can set up and run a (dense network) simulation, making use of already supported functionalities.
Additionally, it is important that users can extend simulators and develop new features to study previously unsupported scenarios. To arrive at an objective assessment of these considerations, we highlight some ease-of-use and extensibility indicators.

\subsubsection{Prerequisite Knowledge}

The prerequisite knowledge necessary to use a simulator greatly affects its ease-of-use and can be a reason to prefer one simulator over another, due to a better understanding of how particular parts of the system are modelled, and due to experience in the programming language or design employed by the simulator. Depending on users' previous experience, the programming language and design of a simulator can simplify usage and understanding.

Regarding the programming skill required for different simulators, this is more a function of the programming language used and the design patterns applied than of the algorithmic implementation of the simulator.
When comparing the languages of different simulators, interpreted languages such as Python and MATLAB are often easier to use than compiled languages such as C and C++, due to dynamic typing, automatic memory management and other benefits; however there is a performance trade-off.

\subsubsection{Documentation}

Without proper documentation, learning how to use a new tool can be a daunting task. Documentation can be provided in the form of explanatory publications containing system block diagrams and listings of exposed parameters, through user guides and tutorials aimed at aiding the user to set up, configure and examine their simulations, or through interactive websites aimed at explaining the details and reasoning behind each simulator feature.
Documentation generators such as Doxygen or Javadoc can also be used to create documentation automatically from source code comments. This enables users to check the purpose of classes and functions and simplifies code navigation.

\subsubsection{User Community and Active Forum}
\label{subsect:user_community}

A helpful user community with an active forum can accelerate simulator familiarisation and clarify documentation ambiguities. The size of the community greatly influences the level of feedback. A large community increases confidence in underlying simulator functionality, as problems are more likely to be discovered.

Open access to simulator source code allows the user community to correct errors and extend functionality. The ability of community members to independently distribute these extensions (open source) can greatly accelerate the rate at which new functionality develops and enables faster correction of inconsistencies.

\subsubsection{Network Deployment Scenarios}\label{subsect:network_deployment_scenarios}

Simulation setup can be accelerated significantly depending on whether simulators support desired scenarios already. Standardized scenarios such as those defined by the \ac{3gpp} allow easy verification, isolation of unique features and enable fair comparison of algorithms, protocols, and technologies for \acp{dsn}. It is also important to consider whether a simulator supports the level of accuracy required for the scenarios chosen.

Ideally, simulators should facilitate easy extension of predefined scenarios, and the development of new scenarios; and be implemented in such a manner that parameter changes do not result in incompatibilities.

\subsubsection{Structure and Modularity}

When designing a complex system, a key concept that the designer must bear in mind is its structure and modularity. By isolating modules according to their functionality, basic components can be made re-usable and their maintenance made easier. Modularity also enables simulator extension by facilitating the development of new components.

As \acp{dsn} become more complex, the concept of modularity becomes increasingly important. In these scenarios, a single node may have multiple radio access technology interfaces, and the ability of a node to switch transmissions between one technology and another is facilitated by a modular design.


\section{Comparative Analysis}

In this section, we compare three simulators regarding their suitability to \aclp{dsn} from technical and operational perspectives. Furthermore, we provide a numerical comparison of the scalability of these simulators to \ac{dsn} scenarios.

\subsection{Representative simulators}\label{subsect:Ownership}

For this comparison we have chosen three simulators, the \textit{Vienna \ac{sl} simulator} for \ac{lte} downlink \cite{Ikuno2010}, the \ac{nl} \textit{ns-3} \ac{lte} module \cite{Piro2011} and an in-house-built \ac{sl} simulator, which we will refer to as \textit{HetDenSim}. All three simulators are designed for \ac{lte}, which we consider to be a good baseline for \ac{dsn} simulations. No \ac{ll} simulator was selected, as \acp{dsn} focus on large network -- rather than single link -- interactions.

HetDenSim targets \ac{hetnets} and large networks of small and macro cells. It was designed with time and frequency domain based \ac{rrm} techniques in mind and is particularly suited to coverage, \ac{sinr} statistics, and expected throughput investigations in dense networks.
HetDenSim's populous small cell network design emphasis makes it a well-suited tool for \ac{dsn} simulations.

The Vienna \ac{sl} simulator, which is freely licensed for academic use, has been designed to assess both Homogeneous and Heterogeneous \ac{lte} network deployments. The Vienna \ac{sl} simulator implements the \ac{phy} with a low level of abstraction and therefore, from the \ac{dsn} perspective, provides a powerful means of testing techniques which require detailed \ac{phy} implementation, such as \ac{mimo}.

ns-3 is an open source general purpose \ac{nl} simulator for both wired and wireless networks; we will mainly refer to the \ac{lte} module for ns-3 \cite{Piro2011}. Given its extensible nature, ns-3 can provide a viable \ac{dsn} simulation environment.


\subsection{Technical Comparison}\label{subsect:technical}

In this section we compare the chosen simulators in terms of the required technical features we have identified in Section \ref{subsect:technical_features}; this comparison is summarized in Table \ref{table:comparison}.

\begin{table*}[tph]
    \caption{Comparative Analysis}
    \label{table:comparison}
    \centering
    \resizebox{1.8\columnwidth}{!}{
    \begin{tabular}{>{\raggedright\arraybackslash}m{0.3\columnwidth}>{\raggedright\arraybackslash}m{0.57\columnwidth}>{\raggedright\arraybackslash}m{0.66\columnwidth}>{\raggedright\arraybackslash}m{0.47\columnwidth}}

    \toprule
	\textbf{Technical Feature} & \textbf{Vienna SL}   & \textbf{ns-3}  & \textbf{HetDenSim}  \\


        \midrule
        Link-level abstraction					& {\ac{miesm}, maps effective \ac{sinr} to \ac{bler}, from curves obtained from \ac{ll} simulator \cite{Mehlfuhrer2011a}}	& {\ac{miesm}, maps effective \ac{sinr} to \ac{bler}, from curves obtained from \ac{ll} simulator \cite{Mehlfuhrer2011a}}& Modified Shannon capacity \cite{Mogensen2007} for \acp{ue}' data rates \\
       \vspace{2mm} Interference Suppressing Receivers		& \vspace{2mm} Only \ac{zf}; however additional linear types can be created. \ac{mimo}	transmissions form a central design feature of this simulator in general			& \vspace{2mm} Linear and non-linear receivers implemented as sets of predefined modifications to the SINR curves to reduce complexity\cite{Catreux2003} &  \vspace{2mm} Not supported \\
     \vspace{2mm} Frequency Reuse				& \vspace{2mm}	Hard frequency reuse and \ac{ffr} for macrocell in hexagonal grid					& 		\vspace{2mm}	Hard, strict, soft, soft fractional, enhanced and distributed fractional frequency reuse  		& \vspace{2mm} Hard frequency reuse and \ac{ffr} for macrocell in hexagonal grid and Frequency Aloha \cite{Chandrasekhar2009} for small cells \\
          \vspace{2mm}Channel model	(fast fading)			& \vspace{2mm}	 ITU Pedestrian A,	ITU Pedestrian B, ITU Extended Pedestrian B, ITU Typical Urban, ITU Vehicular A, ITU Vehicular B, Winner II+-based	& 		\vspace{2mm}ITU Pedestrian A,	ITU Pedestrian B, ITU Extended Pedestrian B, ITU Typical Urban, ITU Vehicular A, ITU Vehicular B 		&\vspace{2mm} Not supported \\
     \vspace{2mm}   Traffic Types							&  \vspace{2mm} Supports full buffer, \ac{voip}, video and several others but models all inter-cell interference as full buffer traffic				& \vspace{2mm} Full buffer, constant bit rate, voice traffic, trace-based video traffic, can be extended to additional traffic models easily  & \vspace{2mm} Full buffer only\\
\vspace{2mm}        Scheduling								&		\vspace{2mm}Full buffer traffic scheduling (\ac{rr} and Channel-aware such as \ac{pf}, \ac{bcqi}, and Max-Min throughput) 			& \vspace{2mm}Full buffer traffic scheduling (\ac{rr}, \ac{pf}, \ac{bcqi}, and Max-Min throughput)	and Channel and QoS-aware and Priority Set scheduling	for non-full buffer traffic		& \vspace{2mm}Full buffer traffic scheduling (\ac{rr}, \ac{pf}, and \ac{bcqi})  \\
      \vspace{2mm}  Upper Layers	 		& \vspace{2mm}Outside simulator scope					& 	\vspace{2mm}\ac{rlc}, \ac{pdcp}, \ac{rrc} and other core network protocols when operated with the \ac{epc}		& \vspace{2mm}Outside simulator scope \\
     \vspace{2mm}   Backhaul								& 			\vspace{2mm} Outside simulator scope			&		\vspace{2mm} Simulate \ac{epc} and
additional transport technologies such as point-to-point links or \acp{pon} \cite{Pedro2013}			&  \vspace{2mm} Outside simulator scope \\
        \vspace{2mm} Emulation								& 	\vspace{2mm}Not considered in design		& 	 \vspace{2mm}Considered in design	& \vspace{2mm}Not considered in design \\
        \midrule
        \textbf{Operational Feature}			&   					&   				&  \\
        \midrule
        \vspace{1mm}Coding Language 						&   MATLAB              &   C++       & MATLAB \\
        \vspace{1mm} No. of primary publication citations\footnotemark[1]                        &   243 \cite{Ikuno2010}        &   39 \cite{Piro2011}           &  5  \\
        \vspace{1mm}No. of forum posts within a month\footnotemark[2]&  28                  &   114           &   No forum   \\
        \vspace{1mm}Licensing   & Free of charge for academic use   & Open Source (GPL) & Proprietary  \\
        \vspace{1mm}Automatic code testing scripts                  &   Limited    &   Yes   &   Limited \\
        \vspace{1mm}Automatic documentation generation              &   No    &   Doxygen   &   No \\
        \bottomrule
    \end{tabular}
    }
\end{table*}

The Vienna \ac{sl} simulator has been designed to simulate the \ac{ran} of LTE and is particularly suited to assess physical layer techniques for interference suppression/mitigation from a system level perspective.
In fact, the physical layer abstraction of the Vienna simulator and the \ac{sinr} computation aid the implementation
of additional linear interference suppressing receiver types, since the
signal, interference and noise components of the \ac{sinr} can all
be exposed individually and the receive filter response can be
applied to each to obtain the post-equalisation \ac{sinr}.

The \ac{lte} module for the ns-3 simulator has been designed with the purpose of evaluating radio level performance and end-to-end \ac{qos} of \ac{lte} networks. Due to its modularity and event-driven nature, ns-3 is suited to study a wide range scenarios. For example, it is easy to implement non-full buffer traffic models, making it a good tool to study \ac{qos}-aware scheduling algorithms. Also, due to the implementation of diverse backhaul technologies and core protocols of \ac{lte}, it can be used to investigate \ac{icic} mechanisms without the assumption of perfect backhaul, handover performance and optical-wireless integration issues. Furthermore, it is possible to emulate the \ac{epc} using this simulator, due to the ns-3 emulation capabilities \cite{gupta2014}.

As opposed to the Vienna \ac{sl} simulator and ns-3, HetDenSim has been designed with the purpose of reducing the required simulation time, specifically for large and dense networks. To achieve this, HetDenSim implements a physical layer model with a higher level of abstraction (i.e., Shannon capacity formula). This speeds up large network simulations at the cost of limitations to algorithms and techniques that require detailed \ac{phy} abstraction.

\subsection{Operational Comparison}\label{subsect:operational}

As shown in table \ref{table:comparison}, the Vienna \ac{sl} simulator and HetDenSim are programmed in
MATLAB, while ns-3 is designed using C++.
\footnotetext[1]{From \url{scholar.google.com} 30/10/2014.}
\footnotetext[2]{For the month 30/09/2014 to 30/10/2014. For Vienna \ac{sl} all forum posts of the thread ``System Level Questions" between these dates are included. As ns-3 is broader than just \ac{lte} only posts containing the keywords ``LTE''  or ``LENA''  are included.} Users with limited programming experience may find Vienna \ac{sl} and HetDenSim faster to learn because of this. However, this comes at the cost of processing speed. C++ usage is often regarded as more complex, since users need to deal with a number of details that MATLAB handles transparently, e.g. memory management. Nonetheless, ns-3 makes use of many advanced design patterns, such as smart pointers, object factories, and functors \cite{Alexandrescu2001}.

Documentation is provided for the Vienna \ac{sl} simulator as explanatory publications (e.g. \cite{Ikuno2010}). For the ns-3 \ac{lte} module, besides \cite{Piro2011}, additional documentation is available in the form of a model guide, which explains the details and reasoning behind each simulator feature, and a user guide to aid users in running simulations. Extra documentation for ns-3 is automatically generated from source code comments using Doxygen. This enables users to check the purpose of classes and functions and simplifies code navigation.

To quantify the community size and activity, some key metrics for the  Vienna \ac{sl} simulator
\cite{Ikuno2010} and the \ac{lte} module for ns-3 \cite{Piro2011} are outlined in the table. HetDenSim does not have a forum due to its proprietary nature. Interestingly, while the main Vienna \ac{sl} publication is more cited, the ns-3 \ac{lte} module forum receives much more activity.

The Vienna \ac{sl} simulator supports a wide range of predefined scenarios, mostly in keeping with the \ac{3gpp} standards. It is possible to define other scenarios, although use of non-default parameter combinations can often result in incompatibilities. The \ac{nl} simulator ns-3 offers a higher level of flexibility. Some predefined scenarios exist, but it is easy to create new scenarios and develop or replace elements of scenarios. HetDenSim has been designed with \ac{dsn} scenarios in mind, and though it is possible to simulate other scenarios, it might be necessary to create the features required for new scenarios.

The illustrative simulators chosen are designed in a modular and \ac{oop} fashion. Vienna \ac{sl} and HetDenSim are focused on \ac{lte}, making it difficult to support other technologies. On the other hand, ns-3 was developed as a general purpose simulator and thus naturally supports other technologies, such as 802.11.

\begin{table}[tbph]
\caption{Simulation Parameters and Hardware Configuration}
\label{table:simulation-parameters}

\centering%
\resizebox{0.95\columnwidth}{!}{
\begin{tabular}{>{\raggedright\arraybackslash}m{0.45\columnwidth}>{\centering\arraybackslash}m{0.55\columnwidth}}
\toprule
\textbf{Parameter}  & \textbf{Value}\tabularnewline
\midrule
 Macro \ac{enb} deployment  & 57 macro \ac{enb} hexagonal grid, 3GPP case 1 \cite[Table A.2.1.1-1]{3GPP36814} \tabularnewline
Pico \ac{enb} deployment  & \vspace{1mm}Outdoor RRH/Hotzone, 3GPP case 6.2 \cite[Table A.2.1.1.2-2]{3GPP36814}, \{0,\,1,\,2,\,3,\,4,\,5\} pico \acp{enb} per macro \ac{enb}\tabularnewline
\ac{ue} deployment &  \vspace{1mm}3420 UEs uniformly distributed\tabularnewline
Propagation model & \vspace{1mm}3GPP Model 1 \cite[Table A.2.1.1.2-3]{3GPP36814}\tabularnewline
Fast fading\footnotemark[3] & \vspace{1mm}Rayleigh Fading (Pedestrian B)\tabularnewline
\ac{ue} Speed &  \vspace{1mm}3 Km/h\tabularnewline
Macro \ac{enb} antenna pattern/gain &  \vspace{1mm}Directional antenna, 3GPP model \cite[Table A.2.1.1-2]{3GPP36814} / 14 dBi\tabularnewline
\vspace{1mm}Pico \ac{enb} antenna pattern/ gain & Omnidirectional / 5 dBi\tabularnewline
\vspace{1mm}\ac{ue} antenna pattern/ gain & Isotropic / 0 dBi\tabularnewline
\vspace{1mm}Bandwidth & 10 MHz\tabularnewline
\vspace{1mm}Carrier Frequency & 2.1 GHz\tabularnewline
\vspace{1mm}Frequency reuse & Full reuse 1\tabularnewline
\vspace{1mm}Traffic & Full buffer\tabularnewline
\vspace{1mm}Scheduler & Round Robin\tabularnewline
\vspace{1mm}Simulation length & 250 TTIs\tabularnewline
\midrule
\textbf{Hardware}  & \textbf{Value}         \tabularnewline
\midrule
Processor   & GenuineIntel i7-4930K CPU@3.4GHz [Family 6, Model 62, Stepping 4]         \tabularnewline
RAM         & 16 GB \tabularnewline
\bottomrule

\end{tabular}}
\vspace{-5mm}
\end{table}

\footnotetext[3]{Fast fading is not performed by HetDenSim.}

\subsection{Scalability for DSNs}\label{subsect:Scalability_for_EDN}
To assess the suitability of the selected simulators for large scale network simulation, this section tests how simulation times scale with
network size.
We investigate the \ac{3gpp} system level scenario for simulation of heterogeneous deployments \cite{3GPP36814}. The main parameters are
reported in Table \ref{table:simulation-parameters}.

\begin{figure}
    \centering
    \includegraphics[width=0.85\columnwidth]{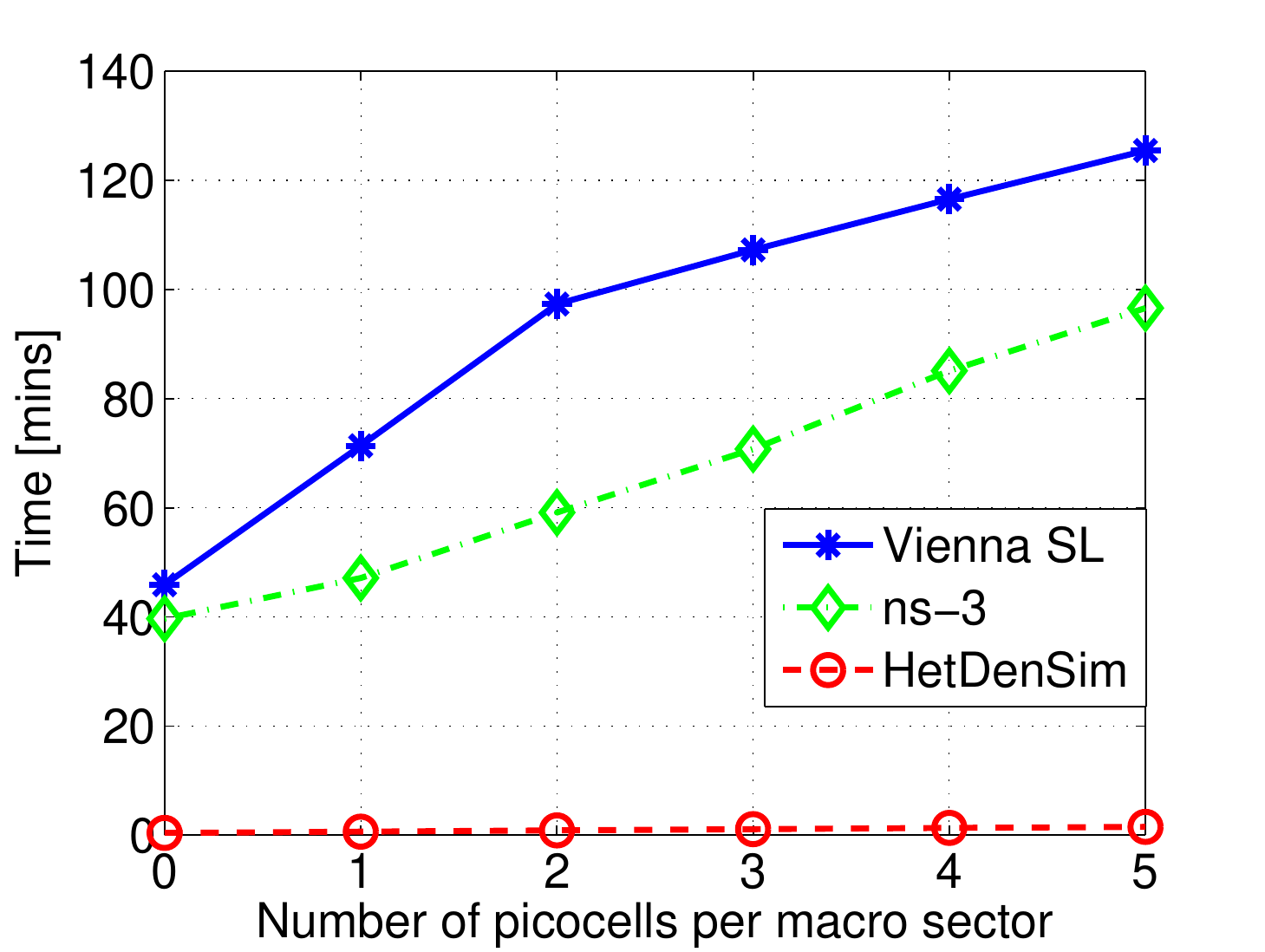}
    \caption{\footnotesize How simulation time varies with small cell deployment density.}
    \label{fig:edn_scaling}
    \vspace{-4mm}
\end{figure}

In Fig. \ref{fig:edn_scaling}, we show the run times of each simulator as we increase the density of low power nodes (i.e. picocells) per
macrocell sector. The simulation time of HetDenSim is shown to be between 65 and 100 times less than that of ns-3, and between 85 and 115
less than that of Vienna \ac{sl}. Moreover, HetDenSim's simulation times scale less steeply with network size than those of ns-3 or Vienna
\ac{sl}. This is attributed to the design of HetDenSim, which makes simplifying assumptions on the \ac{l2s} interface, to speed up simulations.

The lengthy simulation times of Vienna \ac{sl} and ns-3 stem mostly from the detailed \ac{l2s} interface, which despite slowing down the
simulations, makes it possible to investigate effects of many \ac{mimo} techniques and fast-fading effects which are not modeled in
HetDenSim. A few points should be noted; firstly, while Vienna SL and HetDenSim only simulate the downlink, ns-3 also incorporates the
uplink, thus increasing its run times. Secondly, for a fair comparison, the \ac{epc} was not enabled in ns-3. The programming language in
which the simulators were written will also significantly affect the simulation duration. The more lightweight C++ (in which ns-3 is
written) is generally faster than MATLAB (used for Vienna \ac{sl} and HetDenSim).

This example demonstrates the tradeoff between simulation speed, ease-of-use, and the range, precision and types of supported
functionalities, when choosing a simulator.
If we target experiments on interference mitigation techniques, requiring an \ac{l2s} interface with a low level of abstraction, then the choice should aim toward an \ac{sl} simulator, such as the Vienna \ac{sl}. Alternatively, if we need to assess \ac{rrm} algorithms for large networks, then simulators with a less intensive \ac{phy} implementation like HetDenSim would be beneficial in terms of shorter simulation time. If
our aim is to investigate end-to-end \ac{qos}, different traffic types, or the impact of the backhaul in the system, an \ac{nl} simulator such as ns-3 is most suitable.


\section{Conclusion} \label{sec:toward_5G}

There is no single answer to the question "How should \acp{dsn} be simulated?". The answer will be determined by a set of required technical functionalities and practical considerations, specific to the investigation itself. To facilitate readers in answering this question for \acp{dsn}, this work has provided targeted discussion of simulator types and desired features, and comparative analysis of some \ac{dsn}-applicable simulators. %


\newpage

\end{document}